\documentclass[a4paper,11pt]{article}

\usepackage{bm}
\usepackage{latexsym}
\usepackage{amssymb}
\usepackage{amsmath}
\usepackage{mathtools}
\usepackage{graphicx}
\usepackage{mathbbol}
\usepackage{braket}
\usepackage{empheq}
\usepackage{cite}
\usepackage[title]{appendix}
\usepackage{here}
\usepackage{pifont}
\usepackage{booktabs}
\usepackage{slashed}
\usepackage{setspace}
\usepackage{hyperref}
\hypersetup{colorlinks=true, allcolors=blue}

\makeatletter
\renewcommand\subparagraph{
    \@startsection {subparagraph}{5}{\z@ }{3.25ex \@plus 1ex
    \@minus .2ex}{-1em}{\normalfont \normalsize \bfseries }}
\makeatother

\numberwithin{equation}{section}

\begin{document}
\pagestyle{empty}

\vspace{-4cm}
\begin{center}
\hfill YITP-25-149 \\
\hfill KEK-TH-2759 \\
\end{center}

\vspace{1.5cm}

\begin{center}

{\bf\LARGE 
{Lepton number violating signals of a parity \\ \vspace{2mm}
symmetric model at $\mu$TRISTAN
}
}

\vspace*{1.5cm}
{\large 
Keisuke Harigaya$^{1,2,3}$, Ryuichiro Kitano$^{4}$
and Ryoto Takai$^{4,5,6}$
} \\
\vspace*{0.5cm}

{\it 
$^1$Department of Physics, The University of Chicago, Chicago, Illinois
60637, USA \\
$^2$Enrico Fermi Institute, Leinweber Institute for Theoretical Physics, and Kavli Institute for Cosmological Physics,
The University of Chicago, Chicago, Illinois 60637, USA \\
$^3$Kavli Institute for the Physics and Mathematics of the Universe, \\
The University of Tokyo, Kashiwa 277-8583, Japan \\
$^4$Yukawa Institute for Theoretical Physics, Kyoto University,
Kyoto 606-8502, Japan \\
$^5$KEK Theory Center, Tsukuba 305-0801, Japan \\
$^6$The Graduate University for Advanced Studies (SOKENDAI), Tsukuba
305-0801, Japan \\
}

\end{center}

\vspace*{1.0cm}

\begin{abstract}
{\normalsize
\noindent
The parity solution to the strong CP problem necessarily extends the Standard Model to include the SU$(2)_{\rm R}$ gauge sector and imposes restrictions on the structure of the Yukawa interactions. In this framework, one can consider an appealing structure of the neutrino sector in which the smallness of the neutrino masses is naturally explained, while lepton number symmetry is substantially violated at the TeV scale. Observation of distinctive lepton number violating signals at collider experiments can therefore be expected, since the rates are not suppressed by the small neutrino masses.  
We study the constraints from neutrinoless double beta decay and discuss the prospects for discovering new TeV-scale particles, such as the $W'$ boson of SU$(2)_{\rm R}$, via lepton number violating processes at a $\mu^+ \mu^+$ collider, $\mu^+ \mu^+ \to W^+ W'^+$. A $\mu^+ \mu^+$ collider with a center-of-mass energy of 10~TeV can probe the $W'$ boson mass up to about 10~TeV through on-shell production, and the reach can extend to $16$~TeV by studying processes involving off-shell $W'$ boson.}
\end{abstract} 


\newpage
\baselineskip=18pt
\setcounter{page}{2}
\pagestyle{plain}

\setcounter{footnote}{0}

\tableofcontents
\noindent\hrulefill


\section{Introduction}

The TeV energy scale is the next target to be directly explored at collider experiments. 
The LHC experiments have already given limits on some new particle masses up to a few TeV, and the next-generation experiments such as FCC~\cite{FCC:2025lpp}, CEPC~\cite{CEPCPhysicsStudyGroup:2022uwl}, or muon colliders~\cite{InternationalMuonCollider:2025sys} can directly or indirectly search for various kinds of new particles with TeV masses. 
A key theoretical question is what to look for. The most direct motivation for the TeV energy scale comes from Higgs physics, where precision measurements of the Higgs couplings give us more information to build up the real model behind the electroweak symmetry breaking.
At the same time, the future colliders can directly produce TeV scale new particles
and such discovery will change the ``Standard'' of the particle physics to 
include a larger structure such as new symmetry to govern elementary particles.

Introducing symmetries has traditionally been a successful way to organize the parameters in particle physics. 
There have been a series of works aiming to make the microscopic physics parity symmetric, which can address many of the mysteries in the Standard Model (SM). In particular, the strong CP phase, which is odd under P and CP, is forbidden by parity, while the phase in the Cabibbo--Kobayashi--Maskawa (CKM) matrix can be accommodated, as phases in Yukawa interactions are allowed by parity. This provides a natural explanation for the hierarchy in the size of the phases~\cite{Beg:1978mt,Mohapatra:1978fy,Kuchimanchi:1995rp,Mohapatra:1995xd,Babu:1988mw,Babu:1989rb,Hall:2018let}.

In the model we consider, the Higgs field in the SM has a parity partner, $H_{\rm R}$, which transforms as a doublet under a gauge group SU$(2)_{\rm R}$, the parity partner of SU$(2)_{\rm L}$ in the SM~\cite{Babu:1988mw,Babu:1989rb,Hall:2018let}. The vacuum expectation value of $H_{\rm R}$ breaks parity spontaneously, and the model reduces to the SM. 
In such a model, the Yukawa interactions in the SM should be effectively dimension-five operators, which can be obtained by integrating out vector-like fermions.
Unlike the model with SU$(2)_{\rm R}$ breaking by triplet scalars, the strong CP phase is indeed suppressed without introducing extra symmetry. SO$(10)$ grand unification~\cite{Hall:2018let,Mimura:2019yfi,Hall:2019qwx,Carrasco-Martinez:2025zus}, baryogenesis~\cite{Dunsky:2020dhn,Harigaya:2022wzt,Carrasco-Martinez:2023nit,Dasgupta:2025uzi}, and dark matter~\cite{Dror:2020jzy,Baldwin:2024bob,Baldwin:2025oqt} within this framework have been studied in the literature.

The neutrino masses in the parity symmetric model can be realized by introducing three generations of neutral Weyl fermions, $S$~\cite{Hall:2018let,Hall:2023vjb}.\footnote{We may explain the observed neutrino mass by the Dirac mass of the right- and left-handed neutrinos without introducing $S$~\cite{Babu:1988yq,Babu:2022ikf}. Such a model predicts three extra nearly massless Weyl fermions that are produced in the early universe by the exchange of $W'$ and $Z'$ and can be probed by future measurements on the cosmic microwave background.}
The $S$ fields have Yukawa interactions with both left- and right-handed leptons, and thus can form Dirac masses with both left- and right-handed neutrinos. 
At this stage, the lepton number symmetry is substantially violated as $S$ mixes with states with opposite lepton numbers, $\nu_{\rm L}$ and $\bar \nu_{\rm R}$. 
Nevertheless, interestingly, there is no neutrino mass generation at this level. One combination of $\nu_{\rm L}$ and $\bar \nu_{\rm R}$ remains exactly massless. 
Finite neutrino masses are generated only after we introduce a small Majorana mass for $S$ and include quantum corrections. Although the structure is similar, this is very different from the inverse seesaw model, where lepton number is violated by a small parameter to suppress the neutrino masses. 
Therefore, even though the neutrino masses are very small, one can expect large experimental signals of lepton number violation. 
This is a unique feature of this model and provides us with a notable example where the size of lepton number violating processes can decouple from the neutrino masses. 

In this paper, we study lepton number violating signals at a $\mu^+ \mu^+$ collider. 
By setting the initial state to have lepton number, there is a unique process, $\mu^+ \mu^+ \to W^+ W'^+$, where $W'$ is the SU$(2)_{\rm R}$ gauge boson. 
The signals of this process are essentially background free. In the signal events, the total collider energy
goes into hadron jets and charged leptons with a large invariant mass from the $W'$, whereas SM background always comes with escaping neutrinos or leptons. 
We find that a 10~TeV $\mu^+ \mu^+$ collider\footnote{While
the original proposal of $\mu$TRISTAN is limited to lower energies,
higher center-of-mass energies are considered here as benchmarks motivated
by broader discussions of future muon collider capabilities.} can search
for the $W'$ boson up to $16$~TeV by using the off-shell production of
$W'$, depending on the mass of the heavy neutrinos. 
We also compare the reach of the collider experiments with the bound from neutrinoless double beta decay. 
Although the sensitivities cannot be compared directly, as each process probes lepton number violation in the muon and electron sectors, respectively, but
both processes can roughly probe the $W'$ boson up to $\mathcal{O}(10)$~TeV. It is therefore important to study both processes to explore the flavor structure of the model. 

The cross section of the process $\mu^+ \mu^+ \to W^+ W'$ is suppressed when the mass of $S$ is much smaller than $m_{W'}$, as the mass is proportional to lepton number breaking couplings. Such a parameter region may be probed by 
the production
of the heavy neutral fermions, $S$,
via $\mu^+ \mu^+ \to \mu^+ W^+ S$.
We find the production
cross section is as large as
${\cal O}(100\text{--}1000)$~ab when
kinematically accessible and $S$ decays into a charged lepton and a $W$ boson, with lepton flavor violation.
Although we do not perform
the detailed analysis
such as the optimization
of the signal to background
ratio,
we expect that
the process provides
a complementary way 
to search for 
new signals of 
the parity symmetric model.


\section{Model}

\subsection{Parity symmetry and the strong CP problem}

To impose parity, the SM gauge group is extended into ${\rm SU}(3)_{\rm C} \times {\rm SU}(2)_{\rm L} \times {\rm SU}(2)_{\rm R} \times {\rm U} (1)_X$, with parity exchanging ${\rm SU}(2)_{\rm L}$ with ${\rm SU}(2)_{\rm R}$. 
We consider a model with the minimal Higgs content, where ${\rm SU}(2)_{\rm R} \times {\rm U} (1)_X$ is broken down to U$(1)_Y$ by the parity partner of the SM Higgs, $H_{\rm R}$, at the scale $v_{\rm R} \equiv \langle H_{\rm R}
\rangle$.
${\rm SU}(2)_{\rm L}\times {\rm U}(1)_Y$ is broken down to the electromagnetic symmetry by the SM Higgs $H_{\rm L}$ at the scale $v_{\rm L} \equiv \langle H_{\rm L}\rangle$.

The gauge charges of the Higgses and fermions are shown in Tab.~\ref{tab:charges}. Here $q$ $(\ell)$ is the SU$(2)_{\rm L}$ doublet quark (leptons) and $\bar{q}$ $(\bar{\ell})$ is its parity partner. After the spontaneous breaking of ${\rm SU}(2)_{\rm R} \times {\rm U} (1)_X$, $\bar{q}$ splits into two fermions with the same gauge quantum numbers as right-handed up and down quarks and $\bar{\ell}$ does into a fermion with the same gauge quantum number as the right-handed electron and a singlet fermion.
The Yukawa interactions $q \bar{q} H_{\rm L}$ and $\ell \bar{\ell} H_{\rm L}$, that would generate
SM mass terms,
are
forbidden by the gauge symmetry.
Instead, their masses arise from dimension-five operators $\ell \bar{\ell}
H_{\rm L} H_{\rm R}$ for leptons, with analogous operators for quarks,
which are assumed to be completed in some ultraviolet theory.
A simple ultraviolet completion introduces Dirac fermion fields,
$E_\alpha = ({\bm 1}, {\bm 1}, {\bm 1})_{-1}$ and $\overline{E}_\alpha =
({\bm 1}, {\bm 1}, {\bm 1})_1$, allowing charged lepton mass terms
$x^e_{\alpha \beta} \ell_\alpha \overline{E}_\beta H_{\rm L}^* +
x^{e*}_{\alpha \beta} \bar{\ell}_\alpha E_\beta H_{\rm R}^* + M^e_\alpha
E_\alpha \overline{E}_\alpha$.
An alternative scheme replaces these Dirac fermions with a bi-doublet
fermion $\Delta = ({\bm 1}, {\bm 2}, {\bm 2})_0$ to generate Yukawa terms
$x^e_{\alpha \beta} \ell_\alpha \Delta_\beta H_{\rm R} + x^{e*}_{\alpha
\beta} \bar{\ell}_\alpha \Delta_\beta H_{\rm L}$. The up and down masses
may be generated by introducing Dirac fermions $U$, $\overline{U}$, $D$, and $\overline{D}$.
Other possibilities are listed in Ref.~\cite{Hall:2018let}.

\begin{table}[t!]
    \centering
    \caption{The charges of the Higgs bosons and the fermions.}
    \begin{tabular}{cccccccccccccc}
        \toprule
        & & $H_{\rm L}$ & $H_{\rm R}$ & $q$ & $\bar{q}$ & $\ell$ &
        $\bar{\ell}$ & $S$ \\
        \midrule
        ${\rm SU}(3)_{\rm C}$ & & ${\bm 1}$ & ${\bm 1}$ & ${\bm 3}$ &
        ${\bm 3}$ & ${\bm 1}$ & ${\bm 1}$ & ${\bm 1}$ \\
        \addlinespace[1mm]
        ${\rm SU}(2)_{\rm L}$ & & ${\bm 2}$ & ${\bm 1}$ & ${\bm 2}$ &
        ${\bm 1}$ & ${\bm 2}$ & ${\bm 1}$ & ${\bm 1}$  \\
        \addlinespace[1mm]
        ${\rm SU}(2)_{\rm R}$ & & ${\bm 1}$ & ${\bm 2}$ & ${\bm 1}$ &
        ${\bm 2}$ & ${\bm 1}$ & ${\bm 2}$ & ${\bm 1}$ \\
        \addlinespace[1mm]
        ${\rm U}(1)_X$ & & $1/2$ & $-1/2$ & $1/6$ & $-1/6$ & $-1/2$ &
        $1/2$ & $0$ \\
        \bottomrule
    \end{tabular}
    \label{tab:charges}
\end{table}

This framework solves the strong CP problem. Parity forbids the $G\Tilde{G}$ term at higher energy scales. Corrections to it after parity symmetry breaking are also suppressed. 
The up quark masses are generated by
\begin{align}
    {\cal L} = -x^u_{\alpha \beta} q_\alpha \overline{U}_\beta H_{\rm L} -x^{u*}_{\alpha \beta} \bar{q}_\alpha U_\beta H_{\rm R} - M^u_{\alpha \beta} U_\alpha \overline{U}_\beta + {\rm h.c.},
\end{align}
with Hermitian $M^u$.
The mass matrix of $u \subset q$, $\bar{u} \subset \bar{q}$, $U$, and $\overline{U}$ is given by
\begin{align}
    \begin{pmatrix}
        u & U
    \end{pmatrix}
    \begin{pmatrix}
        0 & x^{u} v_{\rm L} \\
        x^{u\dag} v_{\rm R} & M^u
    \end{pmatrix}
    \begin{pmatrix}
        \bar{u} \\ \overline{U}
    \end{pmatrix}.
\end{align}
The determinant of the mass matrix is real, and hence the correction to the strong CP from the up quark masses is absent at the leading order. Similarly, the leading-order correction from the down quark masses is also absent. Higher order corrections are found to be small enough~\cite{Hall:2018let,Hisano:2023izx}.
Note that $x^u$ and $M^u$ are in general complex, and the non-zero CKM CP phase is readily explained.

As we will see, $v_{\rm R} \gg v_{\rm L}$ is required. However, the parity symmetric, tree-level potential of $H_{\rm R}$ and $H_{\rm L}$ is
\begin{eqnarray}
    V = \lambda \left( |H_{\rm R}|^4 + |H_{\rm L}|^4  \right) + \lambda' |H_{\rm R}|^2 |H_{\rm L}|^2 - m^2 \left( |H_{\rm R}|^2 + |H_{\rm L}|^2  \right),
\end{eqnarray}
which does not admit a vacuum with $v_{\rm R} \gg v_{\rm L} \neq 0$. There are two ways to obtain a phenomenologically viable vacuum.
In one, $\lambda' \simeq 2 \lambda$, for which the potential is SU$(4)$ symmetric. The Higgs vacuum expectation values are undetermined at the tree-level, and are fixed by quantum corrections from the top quark Yukawa. The SM Higgs is understood as a pseudo-Nambu--Goldstone boson, and its quartic coupling nearly vanishes at the scale $v_{\rm R}$~\cite{Hall:2018let}.
If the running of the quartic coupling is the SM one, $v_{\rm R} = 10^9$--$10^{13}$~GeV is predicted. With extra Yukawa couplings, $v_{\rm R}$ may be lower.
In another one, parity-breaking masses of $H_{\rm R}$ and $H_{\rm L}$ are introduced, which may be generated from a coupling of the Higgsses with a parity breaking sector. A simple model of parity breaking includes pure Yang--Mills theories with $\theta = \pi$~\cite{Witten:1980sp,tHooft:1981bkw,Witten:1998uka,Gaiotto:2017yup,Kitano:2020mfk}.

\subsection{Neutrino mass}

In this paper, we focus on the neutrino sector whose Lagrangian is given by~\cite{Hall:2023vjb}
\begin{equation}
    \mathcal{L}_\nu = -S_\alpha x_{\alpha\beta} \ell_\beta
    H_{\rm L} - S_\alpha x^*_{\alpha\beta} \bar{\ell}_\beta H_{\rm R} -
    \frac{1}{2} M_{S,\alpha} S_\alpha S_\alpha + {\rm h.c.},
\end{equation}
where $M_S = {\rm diag} \, (M_{S,\alpha})$ denotes the Majorana mass matrix
of the singlets, and $x = (x_{\alpha\beta})$ represents the complex Yukawa
coupling matrix.
Here, the index $\beta$ labels the lepton flavors in the
charged-lepton mass basis.
The determinant of the neutral-lepton mass matrix vanishes, leaving the
active neutrinos massless at tree level.
We assume that $M_S \ll x v_{\rm R}$. The smallness of $M_S$ may be understood by a spontaneously broken symmetry; see Appendix~\ref{sec:ms}.
Note that the lepton number is violated by the Yukawa coupling $x_{\alpha \beta}$ even if $M_S=0$. As we discuss in Sec.~\ref{sec:collider}, this leads to sizable lepton number-violating signals at $\mu^+ \mu^+$ colliders without overproducing neutrino masses.

For simplicity, we assume that the three eigenvalues of the coupling matrix $x_{\alpha\beta}$ are the same; we discuss more generic cases in Sec.~\ref{sec:hierarchical}. Then $x_{\alpha \beta}$ is decomposed  as $x_{\alpha \beta} = x \, \delta_{\alpha i} \, (U_{\rm
PMNS}^\dagger)_{i \beta}$, where $x$ is a positive number. The mass terms can be rewritten as 
\begin{equation}
    \mathcal{L}_\nu \supset - x v_{\rm L} \, S'_i \nu'_i - x v_{\rm R} \,
    S'_i N_i - \frac{1}{2} M_{S, i} \, S'_i S'_i + {\rm h.c.},
\end{equation}
by redefinition of neutral lepton fields as $\nu'_i = (U_{\rm
PMNS}^\dagger)_{i \beta} \, \nu_\beta$, $N_i = (U_{\rm PMNS}^{\rm T})_{i
\beta} \, \bar{\nu}_\beta$ and $S'_i = \delta_{i \alpha} \, S_{\alpha}$.
Two Weyl fermions, $N_i$ and $S'_i$, form a pseudo-Dirac fermion with a
common mass $m_N = x v_{\rm R}$ for all $i = 1, 2, 3$.
The mass $x v_{\rm L} \equiv \theta m_N$ is treated as a
perturbative parameter.
Hereafter, the primes on the neutral lepton fields are omitted.

The active neutrinos $\nu$ are massless at the tree-level and acquire their masses radiatively,
\begin{equation}
    m_{\nu, i} \simeq M_{S, i} \, \frac{m_h^2 + 3 \, m_Z^2}{32 \pi^2 \,
    v_{\rm R}^2} \, ,
\end{equation}
in the limit $m_N \gg m_Z$, $m_h$, where $m_Z$ and $m_h$ are the masses
of the $Z$ boson and the SM-like Higgs boson,
respectively~\cite{Hall:2023vjb}.
The unitary matrix $U_{\rm PMNS}$ is identified as the
Pontecorvo--Maki--Nakagawa--Sakata (PMNS) matrix.

\subsection{Current constraints on \texorpdfstring{$v_{\rm R}$}{vR}}

Constraints on the mass of the $W'$ boson are set by searches at the
LHC, yielding $m_{W'} > 6$~TeV through $W' \to
e \nu$~\cite{ATLAS:2019lsy} and $m_{W'} > 4.6$~TeV through $W' \to t
\bar{b}$ at 95\% confidence level~\cite{ATLAS:2023ibb}.%
\footnote{In our setup, the leptonic decay of $W'$ goes into $e$ and $S$
followed by the decay of $S$ into $\ell W$, $\nu Z$, or $\nu h$. Unless $S$ is
light and stable at collider time scale, the search based on $W'\to \ell \nu$ is
not applicable. We may instead search for $W'^+ \to \ell^+ S$ with $S\to \ell^+
W^{-}$. Here two leptons may be of different flavor. After requiring two
same-sign leptons and two jets consistent with a $W$ boson, the search will be
background free~\cite{CMS:2018jxx}. We expect a bound similar to or stronger
than that from $W' \to e \nu$.  
}
The High-Luminosity LHC (HL-LHC) is expected to be sensitive to $W'$
bosons up to $m_{W'} \sim 8$~TeV through $W' \to e \nu$, $\mu \nu$
and $m_{W'} \sim 5$~TeV through $W' \to t \bar{b}$~\cite{CidVidal:2018eel}.
The other right-handed vector boson $Z'$, with mass $m_{Z'} = m_{W'}
/ \cos \theta_{\rm R}$, is also searched for, where the right-handed
mixing angle $\theta_{\rm R}$ is related to the left-handed one as
$\sin^2 \theta_{\rm R} = \tan^2 \theta_{\rm W} =
0.300$~\cite{ParticleDataGroup:2024cfk}.
Current LHC results give $m_{Z'} > 5.15$~TeV~\cite{CMS:2021ctt}, while
HL-LHC is expected to reach $m_{Z'} \sim 6.7$~TeV~\cite{CidVidal:2018eel},
corresponding to $m_{W'} > 4.3$~TeV and $m_{W'} \sim 5.6$~TeV, respectively.

\begin{figure}[t!]
    \centering
    \includegraphics[width=0.7\textwidth]{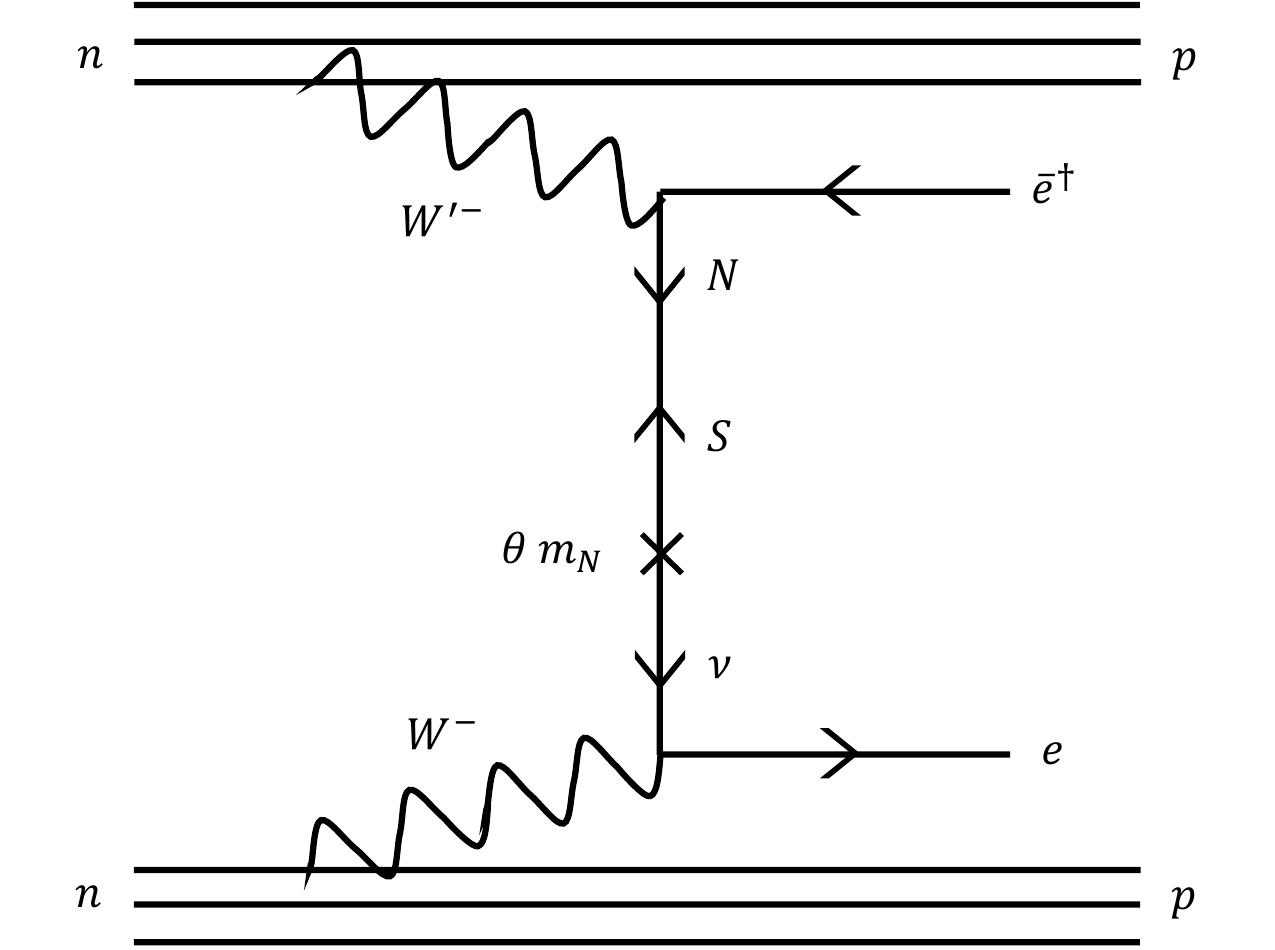}
    \caption{Feynman diagram of the neutrinoless double beta decay
    process, i.e., the $\lambda$-diagram.
    The directions of the fermion arrows are chosen following
    Ref.~\cite{Dreiner:2008tw}.}
    \label{fig:lambda}
\end{figure}

Another probe of lepton number violation involving the $W'$ boson is
neutrinoless double beta decay, mediated by the so-called
$\lambda$-diagram, shown in
Fig.~\ref{fig:lambda}~\cite{Blennow:2010th,Barry:2013xxa}.
The strongest current limit on the half-life is $T^{0 \nu}_{1/2}
\gtrsim 3.8 \times 10^{26}$~yr at 90\% confidence level for
$^{136}$Xe~\cite{KamLAND-Zen:2024eml}.
For the case where the heavy neutral leptons are degenerate,
the half-life is simply expressed by a combination of the PMNS matrix,
$\lambda= U_{\rm PMNS} \, U_{\rm PMNS}^{\rm T}$, as
\begin{equation}
    \left( T^{0 \nu}_{1/2} \right)^{-1} = G^{0 \nu}_{01} \left\vert
    \theta^3 \lambda_{ee} \mathcal{M}_\lambda^{0 \nu} \right\vert^2
    \label{eq:0nubetabeta}
\end{equation}
with $G^{0 \nu}_{01} \simeq 3.78 \times 
10^{-14}$~yr$^{-1}$~\cite{Kotila:2012zza,Mirea:2014dza} and
$\mathcal{M}_\lambda^{0 \nu} = 1.92$--$2.49$~\cite{Borah:2017ldt}.
Interpreting the experimental bound yields $\theta \lesssim 4.7$--$5.2 \times 10^{-3}$ and $m_{W'} \gtrsim 15$--$17$~TeV when we take $\lambda_{ee}$ to be unity.
The absolute value of
$\lambda_{ee}$ can be as small as $0.28$ within $3\sigma$ level of
neutrino mixing parameters~\cite{Esteban:2024eli}, which relaxes the
constraints to $\theta \lesssim 7.2$--$8.0 \times 10^{-3}$ and $m_{W'}
\gtrsim 9.8$--$11$~TeV. 
Meanwhile, the collider signals we consider in the next section 
will not necessarily be suppressed significantly even when $\lambda_{ee}$ is small.
In Sec.~\ref{sec:hierarchical}, we study the case where the three eigenvalues of $x_{\alpha \beta}$ are not equal, for which we find that the bound on $m_{W'}$ from the neutrinoless double beta decay may be further relaxed.


\section{Collider signals}
\label{sec:collider}

\begin{figure}[t!]
    \centering
    \includegraphics[width=0.6\textwidth]{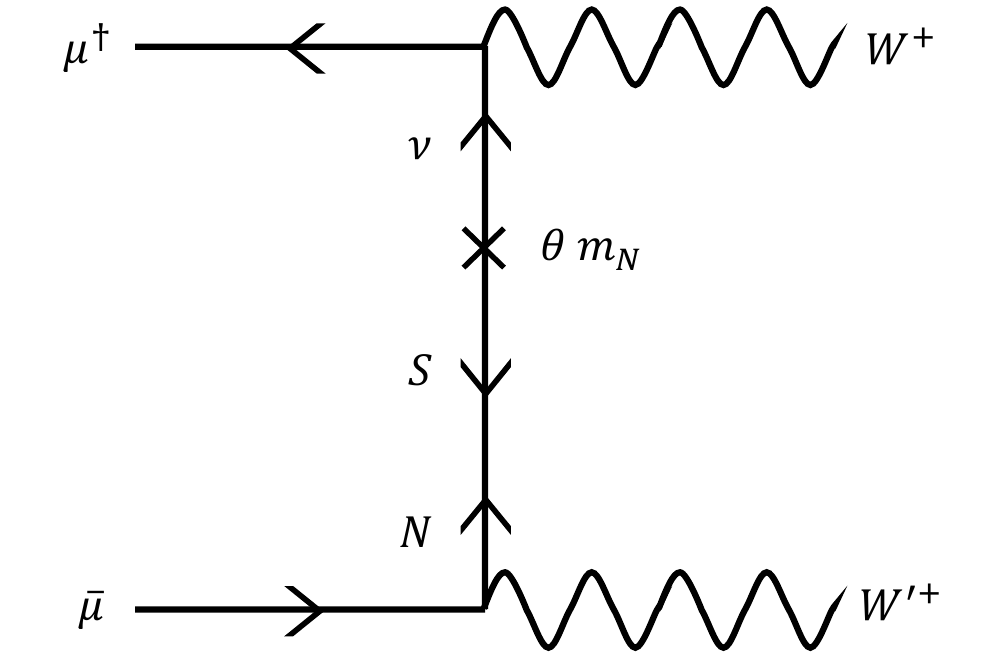}
    \caption{Feynman diagram of the muonic inverse neutrinoless
    double beta decay process, $\mu^+ \mu^+ \to W^+ \, W'^+$.}
    \label{fig:diagram}
\end{figure}

Lepton number violating signals at $\mu^+ \mu^+$ colliders are discussed
in this section.
Our model contains two types of charged vector bosons, $W$ and $W'$,
giving rise to three relevant processes: $\mu^+ \mu^+ \to W^+ W^+$,
$\mu^+ \mu^+ \to W'^+ W'^+$ and $\mu^+ \mu^+ \to W^+ W'^+$.
Though some papers have investigated the first process
(or its charge conjugate)~\cite{Heusch:1995yw,
Raidal:1997tb,Rodejohann:2010jh,Yang:2023ojm,Jiang:2023mte,
Fridell:2023gjx,deLima:2024ohf,Bhattacharya:2025xwv}\footnote{There have also been studies of lepton number violating process at $e^- e^-$ colliders, i.e., $e^- e^- \to W^- W^-$. See Refs.~\cite{Rizzo:1982kn,Gluza:1995ix,Ananthanarayan:1995cn,
Greub:1996ct,Raidal:1997tb}.}, it is suppressed by the small factor $M_S$ in
our model, and so is the second process.
In contrast, the last process, illustrated in Fig.~\ref{fig:diagram},
can occur at high-energy $\mu^+ \mu^+$ colliders.
We assume the use of unpolarized beams in this work, but non-zero polarization is useful in enhancing the signal, suppressing the SM background events, and distinguishing this model with others.
Indeed, our signal is maximized when one antimuon is left-handed and the
other is right-handed. If polarizations of the beams are available, the polarization dependence of the signal will be a non-trivial check of  the nature of lepton number violation.

In the first two subsections, we present the cross sections for muonic
inverse neutrinoless double beta decay, $\mu^+ \mu^+ \to W^+ W'^+$ and
$\mu^+ \mu^+ \to W^+ q q'$ with an off-shell $W'$ boson, focusing on
the case where the heavy neutral leptons are degenerate, $m_{N,i} \equiv
m_N$ for all $i$.
The following subsection investigates single production of heavy neutral leptons as a complementary probe to the lepton number violating processes. Finaly, we illustrates how the model parameters can be chosen without a significant loss of the collider signal while relaxing the constraint from neutrinoless double beta decay.


\subsection{Lepton number violation with \texorpdfstring{$W'$}{W'} boson
production}
\label{sec:on-shell}

First, we consider the case where the center-of-mass energy of the
collider, $\sqrt{s}$, exceeds the $W'$ boson mass, $m_{W'}$.
The mass ratio of the charged vector bosons, $\theta \equiv m_W / m_{W'}
= v_{\rm L} / v_{\rm R}$, is much smaller than unity.
The total cross section of the process $\mu^+ \mu^+ \to W^+ W'^+$ is
\begin{equation}
    \begin{split}
        \sigma (s) &= \frac{\pi \alpha^2 \vert \lambda_{\mu\mu}
        \vert^2}{8 \sin^4 \theta_{\rm W}} \frac{m_N^4}{m_{W'}^4} \left[
        \frac{1}{s} \left( 1 + \frac{2 m_N^2}{s} - \frac{m_{W'}^2}{s}
        \right) \log \frac{s + m_N^2 - m_{W'}^2}{m_N^2} \right. \\
        &\qquad \qquad + \left. \frac{2}{s + m_N^2 - m_{W'}^2} \left( 1
        + \frac{m_N^2}{s} \right) \left( 1 - \frac{m_{W'}^2}{s} \right)
        \left( \frac{m_{W'}^2}{m_N^2} - 1 \right) \right]
    \end{split}
    \label{eq:on-shell}
\end{equation}
at the leading order in $\theta$, where $\alpha = e^2 / 4 \pi$ is the
fine structure constant, $\theta_{\rm W}$ is the (left-handed) weak
mixing angle, and $\lambda_{\mu\mu} = (U_{\rm PMNS} \,
U_{\rm PMNS}^{\rm T})_{\mu\mu}$.
The contribution of the longitudinal $W$ boson dominates over that
of the transverse $W$ boson, since there are 
processes equivalent 
to those with a Goldstone
boson in the final states
when 
the energy scale of the scattering
is significantly higher than the $W$ boson mass.

In the high-energy limit $s \gg m_{W'}^2$, $m_N^2$, the cross section
reduces to
\begin{equation}
    \sigma (s) \simeq \frac{\pi \alpha^2 \vert \lambda_{\mu\mu}
    \vert^2}{8 \sin^4 \theta_{\rm W}} \left[ \frac{m_N^4}{m_{W'}^4}
    \frac{1}{s} \left( \log \frac{s}{m_N^2} - 2\right) +
    \frac{m_N^2}{m_{W'}^2} \frac{2}{s} \right],
\end{equation}
where the first term originates from the emission of longitudinal $W'$
bosons, while the second term corresponds to forward scattering of the
transverse $W'$ boson.

\begin{figure}[t!]
    \centering
    \includegraphics[width=0.8\textwidth]{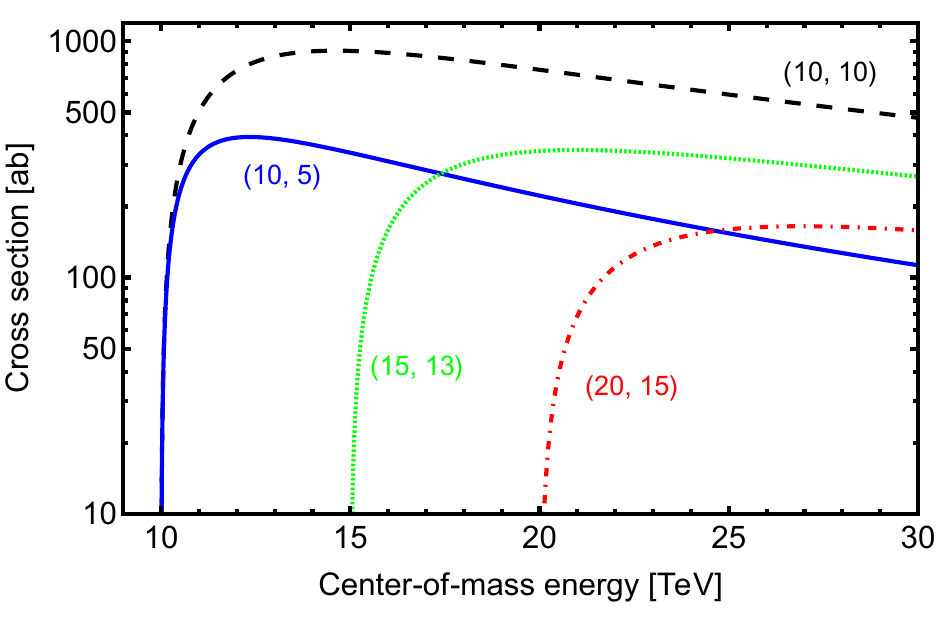}
    \caption{Cross sections of the process $\mu^+ \mu^+ \to W^+ W'^+$
    as functions of the center-of-mass energy, as given in
    Eq.~\eqref{eq:on-shell} with $\lambda_{\mu\mu} = 1$.
    The calculation assumes degenerate heavy neutral leptons and
    unpolarized antimuon beams.
    Each pair of numbers denotes $(m_{W'}, m_N)$ in units of TeV.}
    \label{fig:on_shell}
\end{figure}

We present in Fig.~\ref{fig:on_shell} the total cross section as a
function of the center-of-mass energy of the $\mu^+ \mu^+$ collider
with some parameter choices $(m_{W'}, m_N)$.
For the numerical evaluation, we take the values $\alpha = 1/128$ and
$\sin^2 \theta_{\rm W} = 0.231$~\cite{ParticleDataGroup:2024cfk}, and
assume $\lambda_{\mu\mu} = 1$.
It is observed that the production cross section rises immediately after
the threshold is crossed.

\begin{figure}[t!]
    \centering
    \includegraphics[width=0.9\textwidth]{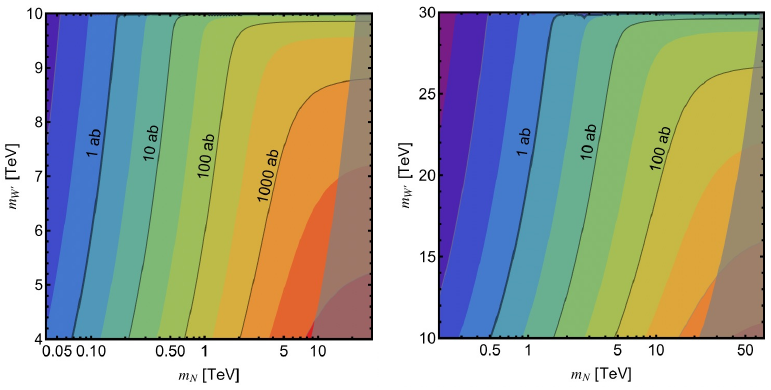}
    \caption{Cross sections of the process $\mu^+ \mu^+ \to W^+ W'^+$
    as functions of $m_N$ and $m_{W'}$, as given in
    Eq.~\eqref{eq:on-shell} with $\lambda_{\mu\mu} = 1$, at the
    center-of-mass energies $\sqrt{s} = 10$~TeV (left) and 30~TeV (right).
    The calculation assumes degenerate heavy neutral leptons and
    unpolarized antimuon beams.
    The shaded region indicates where the perturbativity condition
    $x < 1$ is violated.}
    \label{fig:contour}
\end{figure}

In Fig.~\ref{fig:contour}, we further illustrate the dependence of the
cross section on the heavy neutral lepton mass $m_N$ and the $W'$ boson
mass $m_{W'}$ at the center-of-mass energies $\sqrt{s} = 10$~TeV (left)
and 30~TeV (right).
The shaded region is excluded by perturbativity, $x = g m_N / \sqrt{2}
m_{W'} \lesssim 1$, where $g$ denotes the common ${\rm SU}(2)_{\rm L}$
and ${\rm SU}(2)_{\rm R}$ gauge coupling enforced by parity symmetry.
Over a broad region of the $m_N$-$m_{W'}$ parameter space, one can
expect a cross section of $\mathcal{O}(10)$~ab, corresponding to 100 events at the
integrated luminosity 10~ab$^{-1}$, or more at these collider energies.

The decay width of the $W'$ boson is given by
\begin{equation}
    \Gamma_{W'} \simeq 9 \cdot \frac{\alpha \, m_{W'}}{12 \, \sin^2
    \theta_{\rm W}} + 3 \cdot \frac{\alpha \, m_{W'}}{12 \, \sin^2
    \theta_{\rm W}} \left( 1 - \frac{m_N^2}{m_{W'}^2} \right)^2
    \left( 1 + \frac{m_N^2}{2 m_{W'}^2} \right),
    \label{eq:decay_rate}
\end{equation}
where the second term contributes when $m_N < m_{W'}$.
The first and second terms correspond to hadronic and leptonic decay modes,
respectively, while bosonic decay channels such as $W' \to W \gamma$ and
$W' \to W Z$ are omitted since they do not occur at tree level.
The prefactor in each term reflects the product of the number of colors
and generations for quarks and leptons.
The masses of quarks and charged leptons are ignored.
If the heavy neutral lepton masses are sufficiently close to the $W'$
boson mass, the leptonic decay of the $W'$ boson becomes negligible
compared to its hadronic decay.


\subsection{Lepton number violation via off-shell
\texorpdfstring{$W'$}{W'} boson}

Even when the center-of-mass energy is below the $W'$ boson mass,
lepton-number-violating processes can occur via an off-shell $W'$ boson
decaying into two quarks.
The amplitude can be decomposed as
\begin{eqnarray}
    \mathcal{M} \!\!\! &=& \!\!\! L^\mu (p_1, p_2; p_W, k) \,
    \frac{1}{k^2 - m_{W'}^2} \left( - g_{\mu\nu} + \frac{k_\mu k_\nu}
    {m_{W'}^2} \right) H^\nu (k; p_q, p_{q'}) \nonumber \\
    &=& \!\!\! \frac{1}{k^2 - m_{W'}^2} \sum_{\rm pol.}
    L^\mu (p_1, p_2; p_W, k) \, \epsilon_\mu^* (k) \, \epsilon_\nu (k)
    \, H^\nu (k; p_q, p_{q'}),
\end{eqnarray}
where $L^\mu$ and $H^\mu$ denote the leptonic and hadronic factors,
respectively.
The internal momentum $k$ in the polarization vectors of the $W'$
boson $\epsilon_\mu (k)$ is off-shell, i.e., $k^2 \neq m_{W'}^2$.
Using the recursion relation for the phase space,
\begin{equation}
    {\rm dPS}_3 (P; p_W, p_q, p_{q'}) = {\rm dPS}_2 (P; p_W, k) \,
    {\rm dPS}_2 (k; p_q, p_{q'}) \, \frac{{\rm d} m_{qq'}^2}{2 \pi},
\end{equation}
the cross section of $\mu^+ \mu^+ \to W^+ q q'$ is estimated as
\begin{equation}
    \sigma (s) \simeq \int \frac{{\rm d} m^2_{q q'}}{2 \pi}
    \frac{\hat{\sigma} (s, m_{q q'}^2)}{\big( m^2_{q q'} - m_{W'}^2
    \big)^2 + m^2_{W'} \Gamma_{W'}^2 (m_{q q'})} \, \cdot 2 m_{q q'}
    \, \sum_{q, q'} \Gamma_{W' \to q q'} (m_{q q'}),
\end{equation}
where $m_{q q'}$ is the invariant mass of the final-state quarks,
$\Gamma_{W'} (m) = \Gamma_{W'} \vert_{m_{W'} \to m}$ is the off-shell
decay width of the $W'$ boson, with $\Gamma_{W'}$ defined in
Eq.~\eqref{eq:decay_rate}, and
\begin{equation}
    \begin{split}
        \hat{\sigma} (s, m^2) &= \frac{\pi \alpha^2 \vert
        \lambda_{\mu\mu} \vert^2}{8 \sin^4 \theta_{\rm W}}
        \frac{m_N^4}{m_{W'}^2 m^2} \left[ \frac{1}{s} \left( 1 +
        \frac{2 m_N^2}{s} - \frac{m^2}{s} \right) \log
        \frac{s + m_N^2 - m^2}{m_N^2} \right. \\
        &\qquad \qquad + \left. \frac{2}{s + m_N^2 - m^2} \left( 1
        + \frac{m_N^2}{s} \right) \left( 1 - \frac{m^2}{s} \right)
        \left( \frac{m^2}{m_N^2} - 1 \right) \right]
    \end{split}
\end{equation}
is the modified cross section, Eq.~\eqref{eq:on-shell}, for the
off-shell $W'$ boson.
The factor $1 / m_{W'}^2$ is from $\theta^2$ in the $\nu S$ coupling,
which is not replaced by $1 / m_{qq'}^2$.
Here, polarization correlations are neglected, such that the
polarization-summed factor $\sum_i \vert L^\mu \epsilon_{i \mu}^*
\epsilon_{i \nu} H^\nu \vert^2$ is replaced with $\frac{1}{3} \sum_{i,j}
\vert L^\mu \epsilon_{i \mu}^* \vert^2 \vert \epsilon_{j \nu} H^\nu
\vert^2$~\cite{Uhlemann:2008pm}.

\begin{figure}[t!]
    \centering
    \includegraphics[width=0.8\textwidth]{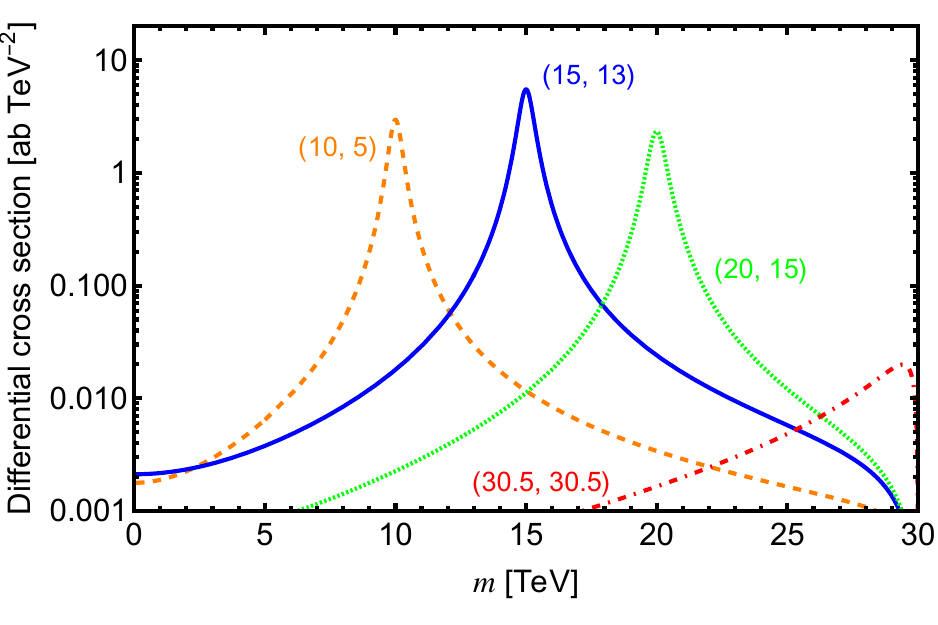}
    \caption{Differential cross sections ${\rm d} \sigma / {\rm d} m^2$
    at $\sqrt{s} = 30$~TeV, shown as functions of the invariant mass
    of the final-state quarks $m$, which corresponds to the integrand in
    Eq.~\eqref{eq:off-shell} with $\lambda_{\mu\mu} = 1$.
    The calculation assumes degenerate heavy neutral leptons and
    unpolarized antimuon beams.
    Each pair of numbers denotes $(m_{W'}, m_N)$ in units of TeV.}
    \label{fig:mDist}
\end{figure}

\begin{figure}[t!]
    \centering
    \includegraphics[width=0.8\textwidth]{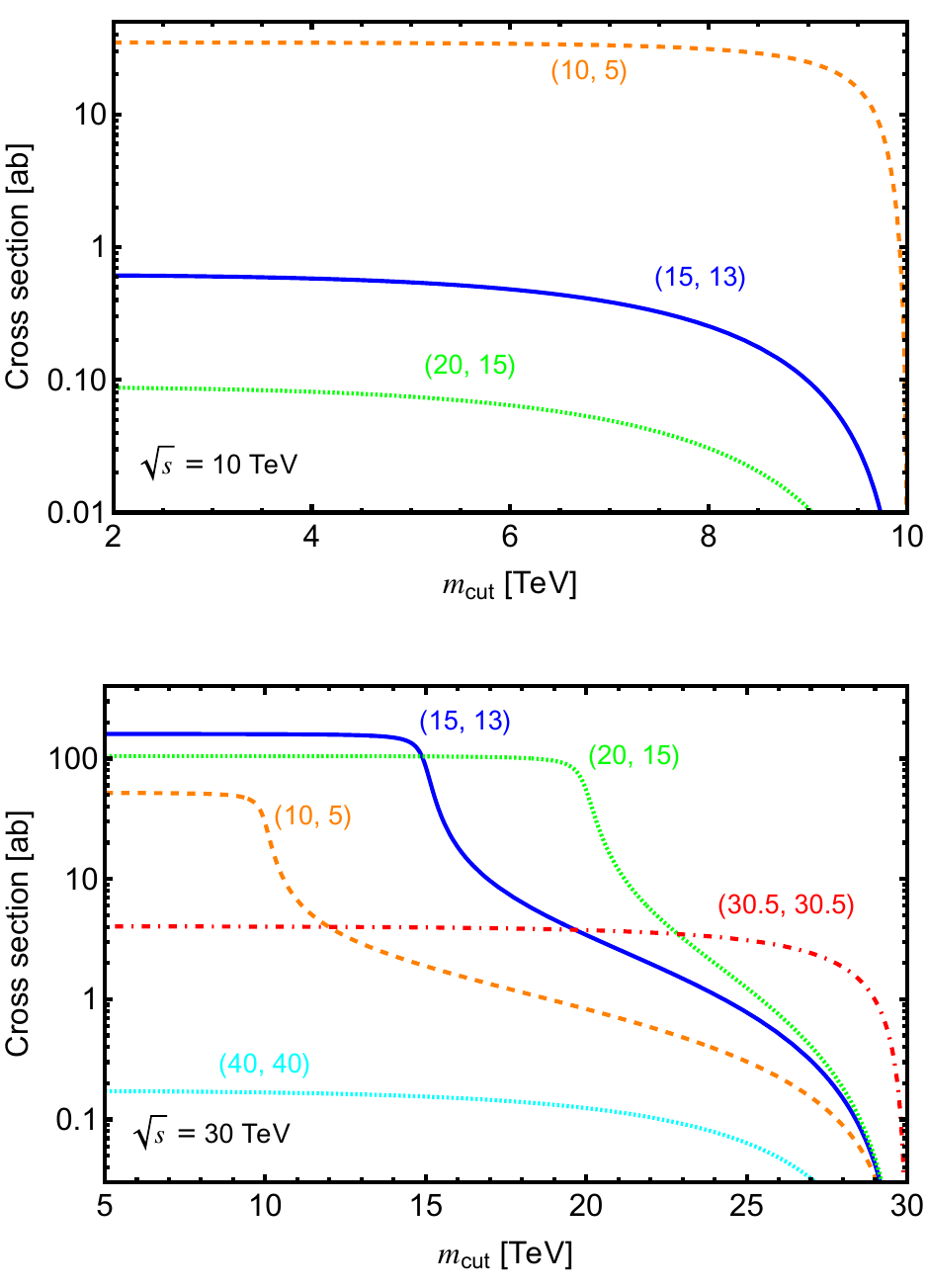}
    \caption{Cross sections of the process $\mu^+ \mu^+ \to W^+ W'^* \to W^+ q q'$ at $\sqrt{s} = 10$~TeV (top) and 30~TeV
    (bottom) ,
    shown as functions of the lower cut on the invariant mass of the
    final-state quarks, $m_{\rm cut}$.
    Each pair of numbers denotes $(m_{W'}, m_N)$ in units of TeV.
    Assumptions include degenerate heavy neutral leptons, unpolarized
    antimuon beams, and $\lambda_{\mu\mu} = 1$.}
    \label{fig:mcutDist}
\end{figure}

Finally, we obtain
\begin{equation}
    \sigma (s) \simeq \frac{3 \, \alpha}{4 \pi \sin^2 \theta_{\rm W}}
    \int^s_{m_{\rm cut}^2} \! {\rm d} m^2 \frac{m^2 \, \hat{\sigma}
    (s, m^2)}{\big( m^2 - m_{W'}^2 \big)^2 + m^2_{W'} \Gamma_{W'}^2 (m)}
    \label{eq:off-shell}
\end{equation}
with a lower cut on the invariant mass of the quarks, $m_{\rm cut}$.
Fig.~\ref{fig:mDist} shows the dependence of the differential cross
section, which corresponds to the integrand in Eq.~\eqref{eq:off-shell},
on the quark invariant mass.
The cross sections at $\sqrt{s} = 10$~TeV and 30~TeV as functions of
$m_{\rm cut}$ are illustrated in Fig.~\ref{fig:mcutDist}.

\begin{figure}[t!]
    \centering
    \includegraphics[width=0.8\textwidth]{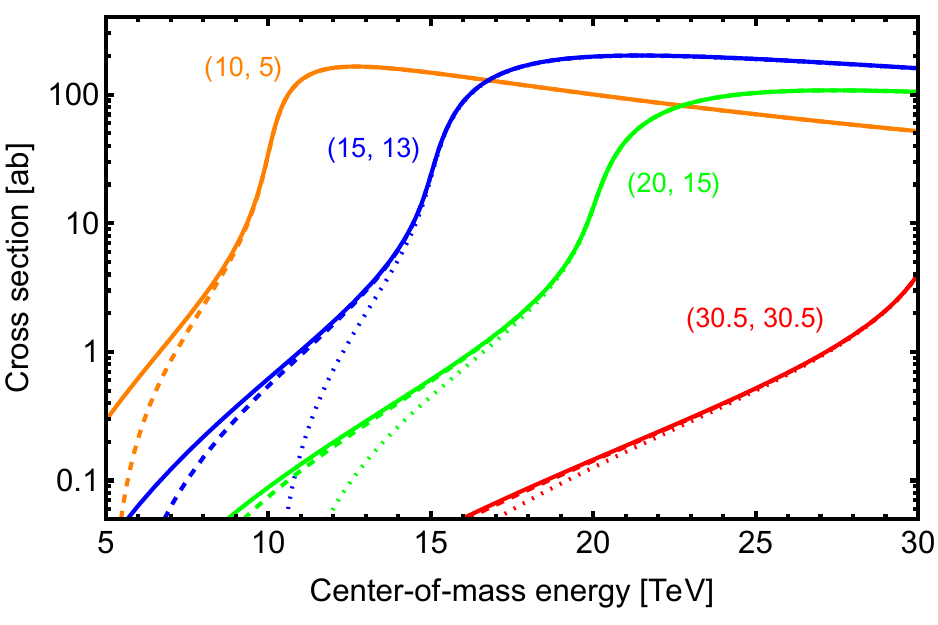}
    \caption{Cross sections of the process $\mu^+ \mu^+ \to W^+ W'^*
    \to W^+ q q'$ as functions of the center-of-mass energy, as given
    in Eq.~\eqref{eq:off-shell}.
    The solid, dashed, and dotted lines indicate results obtained by
    integrating over the quark invariant mass with $m_{\rm cut} = 0$,
    5, and 10~TeV, respectively.
    Each pair of numbers denotes $(m_{W'}, m_N)$ in units of TeV.
    Assumptions include degenerate heavy neutral leptons, unpolarized
    antimuon beams, and $\lambda_{\mu\mu} = 1$.}
    \label{fig:off_shell}
\end{figure}

\begin{figure}[t!]
    \centering
    \includegraphics[width=0.9\textwidth]{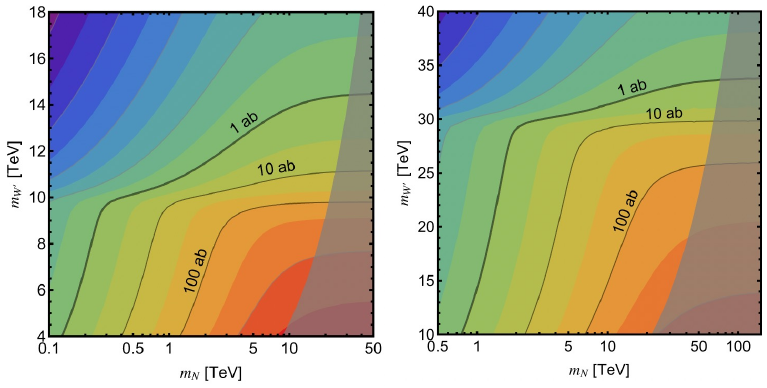}
    \caption{Cross sections of the process $\mu^+ \mu^+ \to W^+ W'^*
    \to W^+ q q'$ as functions of $m_N$ and $m_{W'}$ at the center-of-mass
    energies $\sqrt{s} = 10$~TeV (left) and $30$~TeV (right).
    The calculation assumes degenerate heavy neutral leptons and
    unpolarized antimuon beams, and $\lambda_{\mu\mu} = 1$.
    The shaded region indicates where the perturbativity condition
    $x < 1$ is violated.}
    \label{fig:contour_off}
\end{figure}

Fig.~\ref{fig:off_shell} presents the cross sections as functions of the
center-of-mass energy for several combinations of $(m_{W'}, m_N)$.
The solid, dashed, and dotted lines indicate results obtained by
integrating over the quark invariant mass with $m_{\rm cut} = 0$,
5, and 10~TeV, respectively.
The total cross section at $\sqrt{s} = 10$~TeV and 30~TeV in the
$m_N$-$m_{W'}$ plane are shown in Fig.~\ref{fig:contour_off}.

We discuss below the sensitivity to the lepton number violating signals.
Basically, by imposing a large enough lower cut on the invariant mass of
two quarks in the $W^+ W^{'+*} \to W^+ q q'$ final state, we find that
there is no expected background events while signal events can be
observed. We perform a simple parton level analysis below, which should
be good enough to estimate the sensitivity as the estimated number of
background events are a few order of magnitude smaller than unity.

\begin{figure}[t!]
    \centering
    \includegraphics[width=0.85\textwidth]{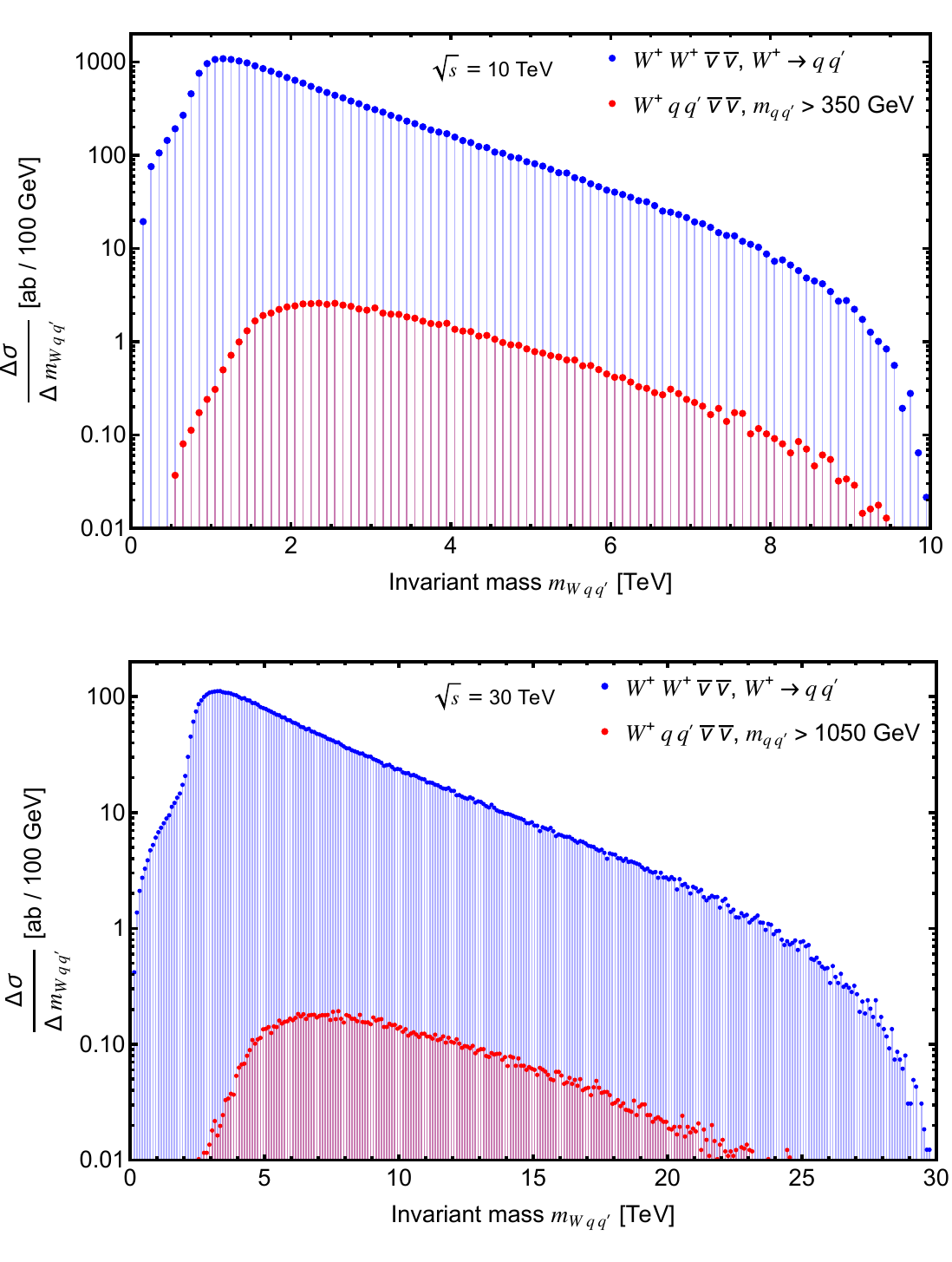}
    \caption{Cross sections estimated in bins of the invariant mass of
    the $W q q'$ system for the Standard Model process $\mu^+ \mu^+ \to
    W^+ q q' \bar{\nu} \bar{\nu}$ with unpolarized antimuon beams at
    center-of-mass energies $\sqrt{s} = 10$~TeV (top) and 30~TeV (bottom).
    Blue points indicate events originating from an on-shell $W$ boson decay
    into two quarks $q$ and $q'$, whereas red points correspond to events
    satisfying the invariant mass cut $m_{q q'} > \sqrt{s} \times 3.5\%$.
    The bin width is set to 100~GeV in both cases.}
    \label{fig:inv_dist}
\end{figure}

For the signal events, as long as $W$ decays hadronically, almost all of the collider energy goes into visible particles. Therefore, the invariant mass of visible particles is peaked at $\sqrt{s}$.
As a reference, we estimate the cross section of a possible SM background 
process $\mu^+ \mu^+ \to W^+ q q' \bar{\nu} \bar{\nu}$ by using
MadGraph~\cite{Alwall:2014hca} at parton level.
Fig.~\ref{fig:inv_dist} shows the distribution of the invariant mass of
the $W$ boson and quarks, $m_{W q q'}$, at $\sqrt{s} = 10$~TeV and
30~TeV, with transverse momentum cuts $p_{\rm T}(W^+)$, $p_{\rm T}(q)$,
$p_{\rm T}(q') > \sqrt{s} \times 3.5\%$.

The blue points show the distribution for $\mu^+ \mu^+ \to W^+ W^+ \bar{\nu} \bar{\nu}$ with at least one of $W^+$ decaying hadronicaly.
One can see that the cross section near $m_{W q q'} \simeq \sqrt{s}$ is suppressed, as the two neutrinos typically carry away a significant portion of the total energy.
Moreover, the $W$ boson would form a single fat jet, and therefore those events can be further reduced by requiring at least three separated jets in the final state.

The red points show the distribution for the case where one of $W^+$ is off-shell and hence three separated jets will exist. Here we further impose $m_{q q'} > \sqrt{s} \times 3.5\%$, which can be implemented by tagging a $W$-jet and taking the invariant mass of two other jets. The resultant cross section near $m_{W q q'} \simeq \sqrt{s}$ is already suppressed, but by imposing stronger cut on $m_{q q'}$, the background can be further suppressed. With a cut $m_{qq'} > \sqrt{s} \times 3.5\%$, the cross section for $m_{W q q'} > \sqrt{s} \times 95\%$ is $1.9 \times 10^{-2}$~ab and $5.4 \times 10^{-3}$~ab for $\sqrt{s} = 10$~TeV and
30~TeV, respectively, implying that no background events are expected
for an integrated luminosity of 10~ab$^{-1}$.

Without background, $\mu^+ \mu^+$ colliders will be able to probe the parameter space with more than 3 expected events. Taking into account the hadronic branching ratio of $W$, this corresponds to $\sigma > 0.5$ ab for an integrated luminosity of 10~ab$^{-1}$ if the efficiency to detect the signal is nearly 100\%.  If $m_N \sim m_{W'}$, $m_{W'} < 16$ TeV can be probed for $\sqrt{s}= 10$ TeV.

In a class of WIMP dark matter models, $m_{W'} < 13$ TeV is required from the relic abundance~\cite{Baldwin:2025oqt}, and $\mu^+ \mu^+$ colliders can probe all of the viable parameter space of such models unless $m_{N} \ll m_{W'}$.


\subsection{Single production of the heavy neutral leptons}

If the heavy neutral leptons are much lighter than the $W'$ boson, the
cross section of the process $\mu^+ \mu^+ \to W^+ W'^+$ is suppressed
by the factors of $m_N^2 / m_{W'}^2$ or $m_N^4 / m_{W'}^4$ as discussed
in Sec.~\ref{sec:on-shell}.
In this case, single production of the heavy neutral leptons,
$\mu^+ \mu^+ \to \mu^+ W^+ S$, becomes important due to the presence of
double large logarithms.

\begin{figure}[t!]
    \centering
    \includegraphics[width=0.8\textwidth]{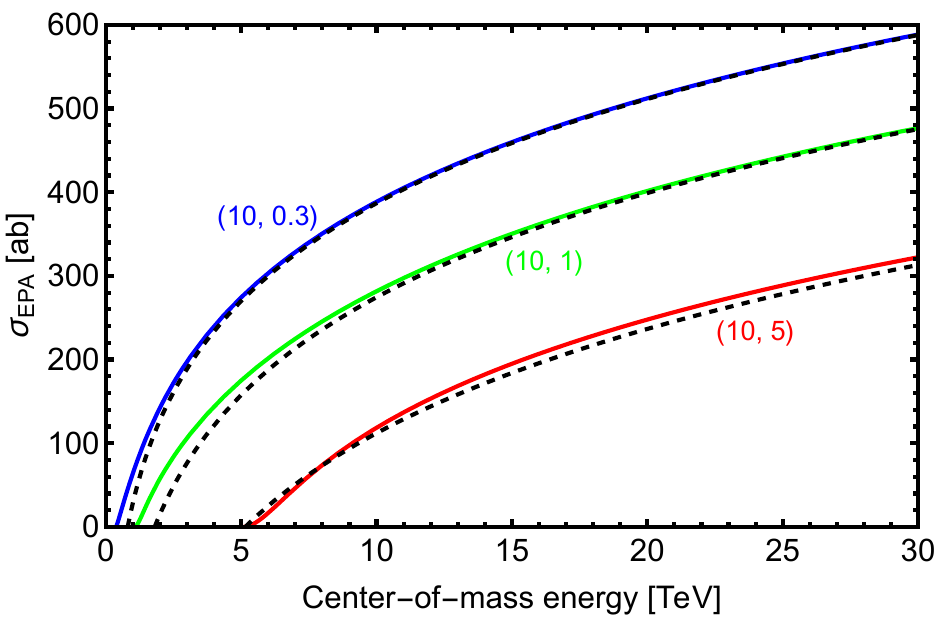}
    \caption{Cross section for the single production of heavy neutral
    leptons in the process $\mu^+ \mu^+ \to \mu^+ W^+ S$, computed using
    the effective photon approximation.
    The solid and dashed lines correspond to the full result calculated
    from Eq.~\eqref{eq:HNL_full} and the high-energy approximation given
    in Eq.~\eqref{eq:HNL_L}, respectively.
    The calculation assumes degenerate heavy neutral leptons and
    unpolarized antimuon beams.
    Each pair of numbers denotes $(m_{W'}, m_N)$ in units of TeV.}
    \label{fig:HNL_EPA}
\end{figure}

\begin{figure}[t!]
    \centering
    \includegraphics[width=0.9\textwidth]{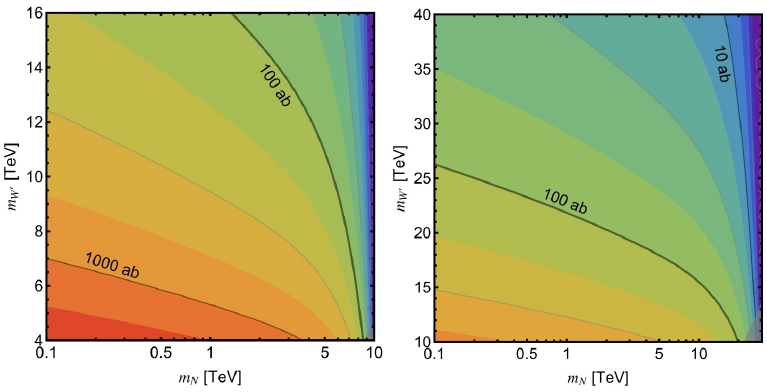}
    \caption{Cross section for the single production of heavy neutral
    leptons in the process $\mu^+ \mu^+ \to \mu^+ W^+ S$, computed using
    the effective photon approximation, at the center-of-mass energies
    $\sqrt{s} = 10$~TeV (left) and 30~TeV (right).
    The calculation assumes degenerate heavy neutral leptons and
    unpolarized antimuon beams.
    The shaded region indicates where the perturbativity condition
    $x < 1$ is violated.}
    \label{fig:hnl}
\end{figure}

The dominant contribution arises from diagrams with collinear photons,
whose representative Feynman diagrams resemble those for single $W$
boson production in the SM, $e^+ e^- \to W^+ \bar{\nu}
e^-$~\cite{Cheyette:1983pk,Philipsen:1992gz}.
In the effective photon approximation, the production cross section is
factorized into the cross section of the subprocess $\mu^+_{\rm R} \,
\gamma \to W^+ S_i$,
\begin{equation}
     \begin{split}
        \sigma_{\mu\gamma} (s) &\simeq \frac{\vert (U_{\rm PMNS})_{\mu i}
        \vert^2 \, \pi \alpha^2}{2 \, m_{W'}^2 \sin^2 \theta_{\rm W}}
        \left( 1 - \frac{m_N^2}{s} \right) \\
        &\quad \times \left[ 1 - \frac{2 \, m_N^2}{s} + \frac{m_N^4}{s^2}
        + \frac{2 \, m_N^4}{s^2} \log \frac{s}{m_W^2} + \frac{4 \,
        m_N^4}{s^2} \log \left( 1 - \frac{m_N^2}{s} \right) \right],
     \end{split}
\end{equation}
and the photon parton distribution function,
\begin{equation}
    f_\gamma (x) = \frac{\alpha}{2 \pi} \left( - \frac{1-x}{x} +
    \frac{1 + (1-x)^2}{x} \log \frac{x s}{m_\mu^2} \right).
    \label{eq:epa_pdf}
\end{equation}
Combining these factors, the cross section via collinear photons is
obtained as
\begin{eqnarray}
    \sigma_{\rm EPA} (s) \!\!\! &=& \!\!\! \int^1_{s_{\rm min} / s}
    {\rm d} x \, f_\gamma (x) \, \sigma_{\mu \gamma} (x s)
    \label{eq:HNL_full} \\
    &\simeq& \!\!\! \frac{\vert (U_{\rm PMNS})_{\mu i} \vert^2 \,
    \alpha^3}{3 \, m_{W'}^2 \sin^2 \theta_{\rm W}} \, \mathsf{L} (s) \,,
    \label{eq:HNL_L}
\end{eqnarray}
where $s_{\rm min} = (m_N + m_W)^2 \simeq m_N^2$ and
\begin{equation}
    \begin{split}
        \mathsf{L} (s) &= \log \frac{s}{m_\mu^2} \log \frac{s}{m_N^2}
        - \frac{1}{4} \left( \log \frac{s}{m_N^2} \right)^{\!\! 2} \\
        &\quad - \frac{103}{24} \log \frac{s}{m_\mu^2} + \frac{1}{2}
        \left( \log \frac{m_N^2}{m_\mu^2} + \frac{1}{3} \right) \log
        \frac{s}{m_W^2} + \frac{9}{4} \log \frac{s}{m_N^2} \,
    \end{split}
\end{equation}
in the limit of $m_W \ll m_N \ll \sqrt{s}$.
The cross section calculated using the effective photon approximation as a
function of the center-of-mass energy is shown in Fig.~\ref{fig:HNL_EPA},
along with its dependence on the $W'$ boson mass and the heavy neutral
lepton mass, presented in Fig.~\ref{fig:hnl}.

We find that when the heavy neutral leptons are at the sub-TeV scale,
the cross section exceeds 100~ab even at 2~TeV colliders, originally
proposed as $\mu$TRISTAN in Ref.~\cite{Hamada:2022mua}.
Furthermore, diagrams involving collinear $Z$ bosons may contribute
subdominantly to the total cross section, adding an $\mathcal{O}(10) \%$
correction, which depends on the beam polarization~\cite{Hamada:2024ojj}.

There are four decay modes of the heavy neutral leptons at tree level,
i.e., $S \to \bar{\nu} h$, $S \to \bar{\nu} Z$, $S \to \ell^+ W^-$,
and $S \to \ell^- q q'$, where $\bar{\nu}$ does not denote the neutral
components of SU$(2)_{\rm R}$ doublets $\bar{\ell}$, but rather
antiparticles of active neutrinos.
The lepton number is conserved in the first three channels, whose
widths are given by
\begin{equation}
    \Gamma (\bar{\nu} h) = \Gamma (\bar{\nu} Z) = \frac{\alpha \,
    m_N^3}{8 m_{W'}^2 \sin^2 \theta_{\rm W}}, \quad  \Gamma (\ell^+ W^-)
    = 2 \vert (U_{\rm PMNS})_{\ell i} \vert^2 \, \Gamma (\bar{\nu} h),
\end{equation}
respectively, in the limit of $m_Z, m_h \ll m_N \ll m_{W'}$.
On the other hand, the width of the last mode that violates the lepton number is
\begin{equation}
    \Gamma (\ell^- q q') = \frac{\vert (U_{\rm PMNS})_{\ell i} \vert^2
    \, \alpha^2 m_N^5}{384 \, \pi m_{W'}^4 \sin^4 \theta_{\rm W}}
\end{equation}
in the same limit.

The signal of the lepton number violation, $S \to \ell^- q q'$,
is very clean due to the lepton charge flip.
However, the branching ratio
\begin{equation}
    {\rm BR} (S_i \to \ell^- q q') \simeq 5.6 \times 10^{-5}
    \times \vert (U_{\rm PMNS})_{\ell i} \vert^2 \frac{m_N^2}{m_{W'}^2}
\end{equation}
would be small for discovery.
The process may be detected via the third decay channel
$S \to \ell^+ W^-$, with a branching ratio of
\begin{equation}
    {\rm BR} (S_i \to \ell^+ W^-) = 0.5
    \times \vert (U_{\rm PMNS})_{\ell i} \vert^2.
\end{equation}
Although lepton number is conserved in the overall process $\mu^+ \mu^+
\to \mu^+ \ell^+ W^+ W^-$, charged lepton flavor is violated for
$\ell = e$, $\tau$.
Also, peaks in the $\ell^+ W^-$ invariant mass distribution may be observable.
In addition, if a large number of antimuons which emit collinear photons can be
detected, the first two modes, whose branching ratios are
${\rm BR} (\bar{\nu} h) = {\rm BR} (\bar{\nu} Z) = 25\%$, cause processes $\mu^+ \mu^+ \to \mu^+ \bar{\nu} W^+ h / Z$, with the
recoil mass of $\mu^+$ and $W^+$ around $m_N$.


\subsection{Non-degenerate case}
\label{sec:hierarchical}

So far we assumed that the three eigenvalues of the Yukawa coupling matrix $x$ are the same. In that case, the neutrinoless double beta decay sets a lower limit on $m_{W'}$ to be about $10$~TeV. 
In this subsection, we discuss the scenario with general Yukawa couplings
$x$ and how the constraint from the neutrinoless double beta decay can be relaxed while the signals at $\mu^+ \mu^+$ colliders can be maintained.

By using the singular value decomposition with two unitary matrices,
$x = U_1 x_{\rm D} V^\dagger$, the mass term can be written as
\begin{equation}
    \mathcal{L}_\nu \supset - S'_i U_{ij} x_j \nu'_j v_{\rm L} - S'_i
    x_i N_i v_{\rm R} - \frac{1}{2} M'_{S, ij} S'_i S'_j + {\rm h.c.},
    \label{eq:mass}
\end{equation}
where we define $U = U_1^{\rm T} U_1$, $S' = U_1^\dagger S$,
$\nu' = V^\dagger \nu$, $N = V^{\rm T} \bar{\nu}$ and $M'_S =
U^{\rm T}_1 M_S U_1$.
This mass term allows us to treat $N_i$ and $S'_i$ as a pseudo-Dirac
fermion with the mass $m_{N,i} = x_i v_{\rm R}$, where $x_{\rm D}
v_{\rm L} \equiv \theta m_N$ is treated as a perturbative parameter.
The active neutrino mass matrix is given by
\begin{equation}
    m_{\nu', ij} = M'_{S, ij} \, \frac{m_h^2 + 3 \, m_Z^2}{32 \pi^2 \,
    v_{\rm R}^2} \, \frac{m_{N,i} \, m_{N,j}}{m_{N,i}^2 - m_{N,j}^2}
    \, \log \frac{m_{N,i}^2}{m_{N,j}^2}
    \label{eq:neutrino_mass}
\end{equation}
in the limit of $m_Z, m_h \ll m_N$~\cite{Hall:2023vjb}.

The scattering amplitudes of $\mu^+ \mu^+ \to W^+ W'^+$ with the
transverse and longitudinal $W'$ bosons are given by
\begin{eqnarray}
    i \mathcal{M}_{{\rm T}, 1} \!\!\! &=& \!\!\! \sum_{i,j} \frac{i g^2
    \, V_{\mu i} \, U^\dagger_{ij} \, V^{\rm T}_{j \mu} \, m_{N,i} \,
    m_{N,j} \sqrt{s} \, t}{\sqrt{2} \, m_{W'} (s - m_{W'}^2)
    (t - m_{N,j}^2)}, \label{eq:amp_t1} \\
    i \mathcal{M}_{{\rm T}, 2} \!\!\! &=& \!\!\! \sum_{i,j} \frac{i g^2
    \, V_{\mu i} \, U^\dagger_{ij} \, V^{\rm T}_{j \mu} \, m_{N,i} \,
    m_{N,j} \sqrt{s} \, u}{\sqrt{2} \, m_{W'} (s - m_{W'}^2)
    (t - m_{N,j}^2)} \label{eq:amp_t2}
\end{eqnarray}
and
\begin{equation}
    i \mathcal{M}_{\rm L} = \sum_{i,j} \frac{i g^2 \, V_{\mu i} \,
    U^\dagger_{ij} \, V^{\rm T}_{j \mu} \, m_{N,i} \, m_{N,j} \, (s +
    m_{W'}^2) \sqrt{t u}}{2 \, m_{W'}^2 (s - m_{W'}^2) (t - m_{N,j}^2)}
    \label{eq:amp_l}
\end{equation}
at the leading order of $\theta$, where $t$ and $u$ denote the
Mandelstam variables.
The theoretical free parameters of the model are the $W'$ boson mass
$m_{W'}$, the heavy neutral lepton masses $m_N$ and the unitary matrix
$U_1$.
Moreover, the PMNS matrix $U_{\rm PMNS} = V U_2^\dagger$, where $U_2$
diagonalizes the neutrino mass matrix $m_{\nu'}$, carries uncertainties
associated with the neutrino mixing parameters~\cite{Esteban:2024eli}.
On the other hand, the half-life of the neutrinoless double beta decay can be
expressed by replacing $\lambda = U_{\rm PMNS} U_{\rm PMNS}^{\rm T}$ in
Eq.~\eqref{eq:0nubetabeta} with $\tilde{\lambda} = V m_N U^\dagger
m_N^{-1} V^{\rm T}$ if the heavy neutral leptons are much heavier than
the typical momentum scale $p \sim 100$~MeV~\cite{Blennow:2010th}.

\begin{figure}[t!]
    \centering
    \includegraphics[width=0.55\textwidth]{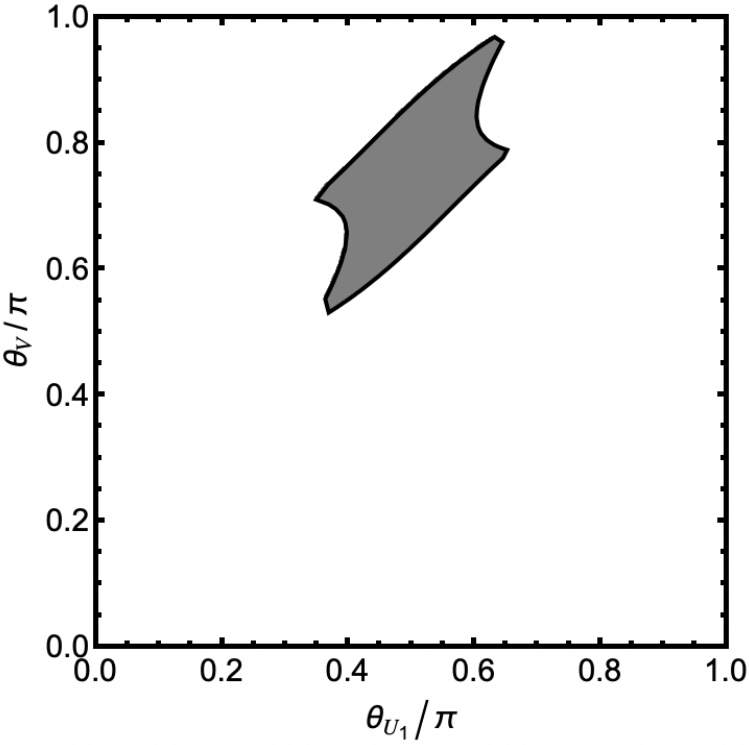}
    \caption{Parameter region of $\tilde \lambda_{ee} < 0.125$ and
    $\tilde \lambda_{\mu \mu} > 0.9$ for the case of $m_{N,1} /
    m_{N,2} = 0.7$ and $m_{N,2} = m_{N,3}$.
    The $x$ and $y$ axes show the rotation angle in $U_1$ and $V$, respectively.}
    \label{fig:tuning}
\end{figure}

In general, it is possible to suppress $\tilde \lambda_{ee}$ while
the signal at the $\mu^+ \mu^+$ collider is kept unsuppressed. Note that the amplitudes in Eqs.~(\ref{eq:amp_t1}--\ref{eq:amp_l}) are proportional to $\tilde \lambda_{\mu \mu}$ when $t \ll m_{N,i}$.
For example, we show in Fig.~\ref{fig:tuning}, the parameter region of $\tilde \lambda_{ee} < 0.125$ and $\tilde \lambda_{\mu \mu} > 0.9$ for the case of 
$m_{N,1} / m_{N,2} = 0.7$ and $m_{N,2} = m_{N,3}$, where the $x$ and $y$ axes show the rotation angle in the following textures of the unitary matrices,
\begin{equation}
    U_1 = \begin{pmatrix}
        \frac{1}{2} (1 + \cos \theta_{U_1}) & \frac{1}{2} (1 - \cos \theta_{U_1}) & \frac{1}{\sqrt{2}} \sin \theta_{U_1} \\
        \frac{i}{2} (1 - \cos \theta_{U_1}) & \frac{i}{2} (1 + \cos \theta_{U_1}) & -\frac{i}{\sqrt{2}} \sin \theta_{U_1} \\
        -\frac{1}{\sqrt{2}} \sin \theta_{U_1} & \frac{1}{\sqrt{2}} \sin \theta_{U_1} & \cos \theta_{U_1}
    \end{pmatrix} \quad \text{and} \quad V = \begin{pmatrix}
        \cos \theta_V & -\sin \theta_V & 0 \\ \sin \theta_V & \cos \theta_V & 0 \\ 0 & 0 & 1
    \end{pmatrix} ,
\end{equation}
respectively. In the region, the lower bound on $m_{W'}$ from the neutrino-less double beta decay is about $8$ TeV and is comparable to the expected sensitivity of the HL-LHC. For $t \gtrsim m_{N,i}$, the signal at the $\mu^+ \mu^+$ collider is not proportional to $\tilde \lambda_{\mu \mu}$, but one can show that similar tuning is possible.

\section{Summary}

Muon colliders are anticipated to directly probe new physics at the
TeV scale by exploiting their high center-of-mass energies and, being
lepton colliders, to simultaneously enable precise measurements of Higgs
and electroweak physics.
In this paper, we demonstrate the potential to search for lepton number
violation at $\mu^+ \mu^+$ colliders with center-of-mass energy
$\mathcal{O} (10)$~TeV, focusing on the process $\mu^+ \mu^+ \to W^+
W'^+$ in the left-right symmetric model ${\rm SU}(3)_{\rm C} \times
{\rm SU} (2)_{\rm L} \times {\rm SU}(2)_{\rm R} \times {\rm U} (1)_X$,
extended with three singlets that form pseudo-Dirac fermions together
with the right-handed neutrino fields. The model features the minimal number of Higgses to break the gauge and parity symmetry while solving the strong CP problem and the minimal number of singlet fermions to explain the non-zero neutrino masses without extra light relics.

We study the scenario in which the Majorana masses of the singlets are suppressed, motivated by an anomaly-free
U$(1)$ symmetry in a certain ultraviolet completion.
In this setup, the lepton symmetry may be violated by unsuppressed Yukawa interactions while the active neutrino mass remains small.
As a result, a lepton number violating process $\mu^+ \mu^+ \to W^+ W'^+$ may be observed at collider experiments.
If the heavy neutral leptons are as heavy as the $W'$ boson, the cross
section can be large enough over a wide parameter space to be
experimentally observable.
Conversely, if their masses lie at the sub-TeV scale, direct production of the heavy neutral leptons would become an important probe.

Constraints from neutrinoless double beta decay bounds 
the model parameter space, particularly implying $m_{W'} \gtrsim 10$~TeV when the heavy neutral leptons are degenerate.
However, if this degeneracy is lifted, parameter tuning can relax the bound while still maintaining a sizable collider signal.
In any case, a high-energy $\mu^+ \mu^+$ collider offers a powerful avenue to explore new physics associated with lepton number violation.


\section*{Acknowledgment}
We thank Liantao Wang for useful discussion.
This work is supported in part by JSPS KAKENHI Grant-in-Aid for
Scientific Research (Nos.~22K21350~[RK]
and 24KJ1157~[RT]),
the U.S.-Japan Science and Technology Cooperation Program in High Energy Physics~(2025-20-2~[RK]),
the U.S.~Department of Energy (DE-SC0009924 [KH]), and World Premier International Research Center Initiative (WPI), MEXT, Japan (Kavli IPMU) [KH].


\appendix
\section{Origin of small Majorana masses}
\label{sec:ms}

In this appendix, we discuss a possible origin of the small Majorana mass $M_S$.

\begin{figure}[t!]
    \centering
    \includegraphics[width=0.8\textwidth]{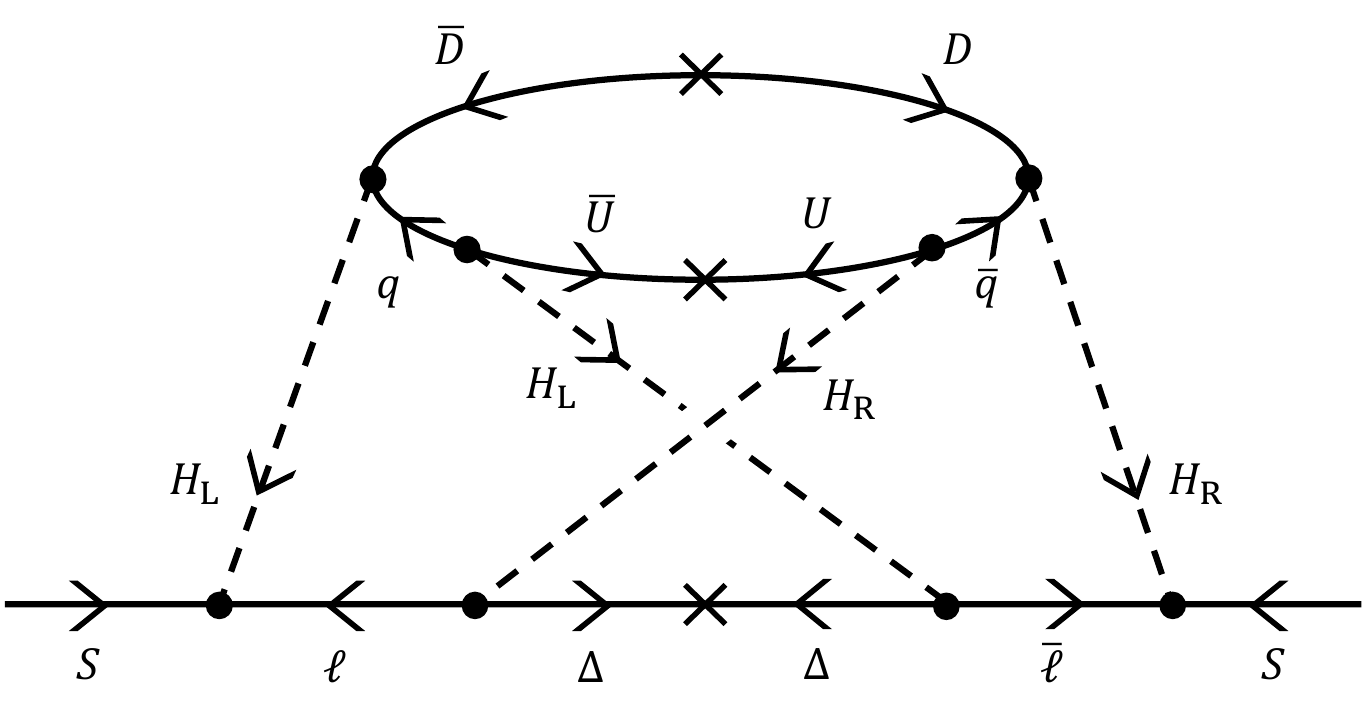}
    \caption{Feynman diagram generating the Majorana mass $M_S$.}
    \label{fig:Ms_loop}
\end{figure}

We first argue that small $M_S$ is quantum mechanically stable.
The Yukawa interactions and masses of the model is
\begin{align}
    x^u \bar{U} q H_{\rm L} + x^{u*} U \bar{q} H_{\rm R} + x^d \bar{D} q H_{\rm L}^* + x^{D*} D \bar{q} H_{\rm R}^* + x^e \Delta \bar{\ell} H_{\rm L}^* + x^{e*} \Delta \ell H_{\rm R}^* + x S \ell H_{\rm L} + x^* S \bar{\ell} H_{\rm R} \nonumber \\ + M_U U \bar{U} + M_D D \bar{D} + \frac{1}{2}M_\Delta \Delta^2+{\rm h.c.},   
\end{align}
where we have suppressed the generation indices.
The leading quantum correction to $M_S$ in the limit $M_S=0$ is given by the 4-loop diagram in Fig.~\ref{fig:Ms_loop} and
\begin{equation}
    \Delta M_S \sim \frac{1}{(16\pi^2)^4} (x^u)^2 (x^d)^2 (x^e)^2 x^2 \frac{M_U M_D M_\Delta}{{\rm max}(M_U^2, M_D^2, M_\Delta^2)},
\end{equation}
where we assume that at least one of $M_U$, $M_D$, and $M_\Delta$ is larger than $v_{\rm R}$, so that the quantum correction is suppressed by the largest of them. Assuming $\mathcal{O}(1)$ Yukawa couplings, the quantum correction is the largest for the third generation. Using $(x^e)^2 v_{\rm R}/ M_\Delta \simeq y_\tau$ and $(x^d)^2 v_{\rm R}/ M_D \simeq y_b$, $M_\Delta$ is the largest and we find
\begin{equation}
    \Delta M_S \sim \frac{1}{(16\pi^2)^4} \frac{y_\tau}{y_b} (x x^u x^d)^2 M_U \sim 10^{-9} \times (x x^u x^d)^2 M_U.
\end{equation}
The corresponding correction to the neutrino mass is
\begin{eqnarray}
 \Delta m_\nu \sim 10~{\rm meV} \times  \frac{10~{\rm TeV}}{v_{\rm R}} (x x^u x^d)^2 \frac{M_U}{v_{\rm R}},
\end{eqnarray}
which can be small enough.

Even if small $M_S$ is quantum mechanically stable, there is a question of why the tree-level $M_S$ is small. The smallness of $M_S$ may be understood by some spontaneously broken symmetry. For example, we may consider a U$(1)$ symmetry shown in the last raw of Tab.~\ref{tab:charges2}.
We assume that the U$(1)$ symmetry is broken by an order parameter ${\cal O}$ with an U$(1)$ charge of 2.
$M_{U,D,\Delta}$ is forbidden by the U$(1)$ symmetry but can be generated by  ${\cal O}U\bar{U}$, ${\cal O}^\dag D\bar{D}$, and ${\cal O}\Delta^2$, while $M_S$ requires ${\cal O}^3 S^2$. Assuming hierarchy between the cutoff scale and the U$(1)$ symmetry breaking scale, tree-level $M_S$ may be suppressed enough.

We may also gauge the U$(1)$ symmetry.
The U$(1)$ symmetry does not have Adler--Bell--Jackiw anomaly, except for U$(1)^3$ and U$(1)\mathchar`-$(gravity)$^2$ anomaly, but they can be canceled by three Weyl fermions with U$(1)$ charges of $-1$.

\begin{table}[t!]
    \centering
    \caption{The charges of the Higgs bosons and the fermions. Here U$(1)$ is an extra symmetry that can explain the smallness of $M_S$.}
    \begin{tabular}{cccccccccccccc}
        \toprule
        & & $H_{\rm L}$ & $H_{\rm R}$ & $q$ & $\bar{q}$ & $\ell$ &
        $\bar{\ell}$ & $U$ & $\overline{U}$ & $D$ & $\overline{D}$ &
        $\Delta$ & $S$ \\
        \midrule
        ${\rm SU}(3)_{\rm C}$ & & ${\bm 1}$ & ${\bm 1}$ & ${\bm 3}$ &
        ${\bm 3}$ & ${\bm 1}$ & ${\bm 1}$ & ${\bm 3}$ & ${\bm 3}$ &
        ${\bm 3}$ & ${\bm 3}$ & ${\bm 1}$ & ${\bm 1}$ \\
        \addlinespace[1mm]
        ${\rm SU}(2)_{\rm L}$ & & ${\bm 2}$ & ${\bm 1}$ & ${\bm 2}$ &
        ${\bm 1}$ & ${\bm 2}$ & ${\bm 1}$ & ${\bm 1}$ & ${\bm 1}$ &
        ${\bm 1}$ & ${\bm 1}$ & ${\bm 2}$ & ${\bm 1}$  \\
        \addlinespace[1mm]
        ${\rm SU}(2)_{\rm R}$ & & ${\bm 1}$ & ${\bm 2}$ & ${\bm 1}$ &
        ${\bm 2}$ & ${\bm 1}$ & ${\bm 2}$ & ${\bm 1}$ & ${\bm 1}$ &
        ${\bm 1}$ & ${\bm 1}$ & ${\bm 2}$ & ${\bm 1}$ \\
        \addlinespace[1mm]
        ${\rm U}(1)_X$ & & $1/2$ & $-1/2$ & $1/6$ & $-1/6$ & $-1/2$ &
        $1/2$ & $3/2$ & $-3/2$ & $-1/3$ & $1/3$ & $0$ & $0$ \\
        \addlinespace[1mm]
        \hline
        \addlinespace[1mm]
        ${\rm U}(1)$ & & $1$ & $1$ & $0$ & $0$ & $2$ & $2$
        & $-1$ & $-1$ & $1$ & $1$ & $-1$ & $-3$ \\
        \bottomrule
    \end{tabular}
    \label{tab:charges2}
\end{table}


\bibliography{bibcollection}
\bibliographystyle{modifiedJHEP}


\end{document}